\documentclass{article}
\usepackage{authblk} 
\usepackage[utf8]{inputenc}
\usepackage[T1]{fontenc}
\usepackage{libertine}
\usepackage{newtxmath}
\usepackage{bm}
\usepackage[a4paper, total={7in, 8in}]{geometry}
\usepackage{graphicx}
\usepackage{hyperref}
\usepackage{multicol}
\usepackage{float}
\usepackage{blindtext}
\usepackage{siunitx}
\DeclareSIUnit{\angstrom}{\text{\AA}}
\usepackage[numbers,square,sort&compress]{natbib}
\usepackage{xcolor}
\usepackage{amsmath}
\usepackage{amsfonts}
\usepackage{mathtools}
\usepackage{nccmath}
\usepackage{verbatim}
\usepackage{longtable}
\usepackage{booktabs}
\usepackage[normalem]{ulem}
\usepackage{textgreek}
\usepackage{textcomp}
\usepackage{dblfloatfix}
\usepackage{placeins}
\usepackage{caption}
\usepackage[version=4]{mhchem} 

\definecolor{darkgreen}{rgb}{0.0, 0.8, 0.4}  
\hypersetup{colorlinks=true,linkcolor=black, filecolor=magenta, citecolor=black, urlcolor=cyan,pdftitle={TwinDomainsinVanDerWaalsQuaternaryOxides},pdfauthor={D.S.Mader}, pdfsubject={SI}, pdfcreator={LaTeX}} 

\newcommand{\wn}{\textrm{cm}^{-1}}
\newcommand{\tsim}{\ensuremath{\sim}}

\title{Twin Domains in Van der Waals Quaternary Oxides}

\author[1]{Dorothée S. Mader}
\author[1,2]{Niels Brumby}
\author[3,4,5]{Xiaosheng Yang}
\author[2,6]{Nele Stetzuhn}
\author[7]{Eduardo Ortega}
\author[8]{Christian Carbogno}
\author[2]{Katayoun Gharagozloo-Hubmann}
\author[1,9,10]{Sebastian F. Maehrlein}
\author[1]{Martin Wolf}
\author[2]{Kirill Bolotin}
\author[3,4,5]{Peining Li}
\author[1,2,$\dag$]{Niclas S. Mueller}
\author[1,$\dag$]{Alexander Paarmann}

\affil[1]{Department of Physical Chemistry, Fritz Haber Institute of the Max Planck Society, Berlin, Germany}
\affil[2]{Department of Physics, Freie Universität Berlin, Berlin, Germany}
\affil[3]{Wuhan National Laboratory for Optoelectronics and School of Optical and Electronic Information, Huazhong University of Science and Technology, Wuhan, China}
\affil[4]{Optics Valley Laboratory, Hubei, China}
\affil[5]{Hubei Optical Fundamental Research Center, Wuhan, China}
\affil[6]{Max Born Institute for Nonlinear Optics and Short-Pulse Spectroscopy, Berlin, Germany}
\affil[7]{Department of Interface Science, Fritz Haber Institute of the Max Planck Society, Berlin, Germany}
\affil[8]{Theory Department, Fritz Haber Institute of the Max Planck Society, Berlin, Germany}
\affil[9]{Institute of Applied Physics, Technische Universität Dresden, Dresden, Germany}
\affil[10]{Institute of Radiation Physics, Helmholz Zentrum Dresden-Rossendorf, Dresden, Germany}
\affil[$\dag$]{Corresponding authors: niclas.mueller@fu-berlin.de, alexander.paarmann@fhi-berlin.mpg.de}

\date{}

\begin{document}
\flushbottom

\maketitle
\begin{abstract}
\label{abstract}
Optical anisotropy is the basis for many intriguing phenomena in van der Waals materials, including hyperbolic polaritons and extreme birefringence. Stacking and twisting van der Waals materials along the out-of-plane direction emerged as a powerful route to tailor this anisotropy, but designing lateral interfaces remains a challenge. Here, twin domains are reported in the van der Waals quaternary oxides \ce{MgTeMoO6}, \ce{MnTeMoO6}, \ce{ZnTeMoO6}, and \ce{CoTeMoO6} – materials that possess strong in-plane optical anisotropy and second-order nonlinearity. The domains naturally form in their orthorhombic crystal structure and extend over hundreds of micrometers. It is proposed that this stability is achieved by the domain wall acting as a diagonal mirror plane in the crystal structure, parallel to the (1-10) or (110) crystal planes, resulting in nearly opposite birefringence between domains. This hypothesis is experimentally confirmed by determining the angle between the crystal axes of neighboring domains using polarization-resolved optical microscopy, infrared-visible sum-frequency generation microscopy, and transmission electron microscopy. The latter further allowed an estimate of the domain wall thickness. 
Overall, the observation of twin domains with orthogonal optical anisotropy opens new routes to use van der Waals quaternary oxides for birefringent waveguiding, polariton steering, and frequency conversion applications.   
\end{abstract}

\begin{multicols}{2}

\section{Introduction} 
\label{intro}
Birefringence is an important physical phenomenon in modern technology as it enables polarization and beam control of light with waveplates, polarizers or filters \cite{ermolaev2021giant}. Van der Waals (vdW) materials, which possess intrinsic birefringence due to their 2D layered structure, enable a miniaturization of such optical elements to the nanoscale \cite{T24_de2025roadmap}. First works on birefringent optical components based on vdW materials made use of black phosphorous, transition metal dichalcogenides, or molybdenum trioxide \cite{ermolaev2021giant,T15_yang2017optical,enders2025mid}, but in-plane anisotropic vdW materials that are also non-toxic \cite{cooper2007vanadium} and stable under ambient conditions \cite{Island_2015,Gao2016TMDCAging} remain scarce. Vertically stacking and twisting vdW materials into heterostructures enables tailoring their optical and electronic properties through modulations of the vertical interlayer interactions \cite{T24_de2025roadmap, Hu2020}. Designing lateral vdW interfaces and heterostructures, on the other hand, remains challenging and typically requires nanofabrication approaches or special growth techniques that are usually limited to monolayers \cite{T24_de2025roadmap, Chakraborty2025}.

Recently, vdW quaternary oxides (QO) with the formula \ce{ATeMoO6}, where A = Cd, Mg, Mn, Co or Zn, attracted attention for their nonlinear optical properties and hyperbolic phonon polaritons
\cite{O17_sun2024van,O18_liu2024air,O20_liu2026regulating,O21_Li2026}. Though already synthesized in the 70-80s
\cite{O1_botto1980x,O2_forzatti1980metal,S1_nonloczynski1976methods,S2_kozlowski1976cotemoo6,S3_forzatti1977synthesisMn,S4_forzatti1977synthesisCd,S5_forzatti1977synthesisZn}, new growth methods and advanced characterization techniques enabled improved crystal quality and detailed understanding of their crystal structure \cite{S8_laligant2001x,S9_doi2009magnetic,S10_doi2009crystal, O4_zhang2012mgtemoo, O5_jin2013top,O6_zhao2013zntemoo}. Similar to most vdW crystals these QOs possess a large out-of-plane anisotropy due to their layered structure \cite{O13_li2018controlled,O15_li2019investigations}. More importantly, biaxial QOs with  A = Mg, Mn, Co or Zn further exhibit a strong in-plane optical anisotropy, which gives rise to in-plane birefringence and in-plane hyperbolicity \cite{O17_sun2024van, O18_liu2024air}. Moreover, the crystal structure of QOs lacks inversion symmetry, giving rise to second-order nonlinearities that are surprisingly large despite their space groups being nonpolar \cite{O4_zhang2012mgtemoo}. This apparent contradiction has been attributed to the second-order Jahn-Teller effect, which leads to a distortion of the coordination environment of the \ce{AO6} building blocks as well as the stereo-active lone pair cation in the \ce{TeO4} building blocks \cite{O4_zhang2012mgtemoo,O6_zhao2013zntemoo,O8_zhao2013combination,O9_jiang2014role,O10_cammarata2014microscopic}.

Apart from their large nonlinear susceptibility, QOs are promising candidates for technological applications in mid-IR laser physics due to their moderate-to-high optical damage thresholds \cite{R10_Guo2022}, high stability in air \cite{O18_liu2024air,R12_Zhou2025}, high chemical stability \cite{R12_Zhou2025}, high optical homogeneity \cite{O13_li2018controlled, R10_Guo2022}, low toxicity compared to vanadium alternatives \cite{R10_Guo2022} and large bandgap leading to a wide transparency region in the mid-IR
\cite{O4_zhang2012mgtemoo, O5_jin2013top, O6_zhao2013zntemoo, O13_li2018controlled, O8_zhao2013combination, R12_Zhou2025, O11_mkaczka2014growth, O14_li2018electric}. The group of X.Tao already tested the related quaternary oxide \ce{BaTeMo2O9} in applications as a Raman laser \cite{A8_Gao2012APL,A9_Gao2013OptExp}, a Q-switch \cite{A10_Wu2017BaTeMo2O9,A12_Liu2021AlphaBaTeMo2O9}, and a polarizing prism \cite{A14_Guo2022BaTeW2O9}. Similar  applications are expected for the QOs discussed in this paper even though most previous work focused on applications in catalysis \cite{A1_forzatti1978solid,A2_forzatti1978cdtemoo6,A3_grzybowska1985co,A4_forzatti1979oxidation,A5_Hayashi1997Telluromolybdates,A6_Hayashi1999}.

While single-crystalline quaternary oxides already show promising optical properties, we now report on twin domains in this vdW materials class.
These domains could solve the challenge of engineering lateral interfaces in vdW materials and open new routes toward nanodevices based on birefringence. Domains in \ce{CoTeMoO6} were previously reported by Li et al. \cite{S24_Li2025}, however the work focused on the magnetic properties of the material. Here, we observe twin domains in the broader set of \ce{ATeMoO6}, with A = Mg, Zn, Mn, and Co, and characterize their structural and optical properties in detail. The domains constitute well-defined in-plane interfaces that extend over hundreds of micrometers. This large spatial extent and the natural occurrence in the orthorhombic crystal structure suggest a high stability of the domain walls (DWs). Our experiments demonstrate that the DWs act as diagonal mirror planes, parallel to the (1-10) or (110) crystal planes, causing nearly opposite birefringence between the domains,\emph{i.e.},  polarization rotation in opposite directions. 

Using polarized optical microscopy (POM) and point spectroscopy, we probe the in-plane birefringence of neighboring domains. The results indicate that the crystal axes are rotated by \tsim~93\textdegree\,at the DWs, leading to opposite in-plane birefringence. This geometry of the domain walls is further confirmed using transmission electron microscopy (TEM). The abrupt changes in in-plane anisotropy are also imprinted onto the nonlinear optical response. Using sum-frequency generation (SFG) microscopy combining mid-infrared (mid-IR) and visible (VIS) lasers, we demonstrate a second-order nonlinear response that is strongly enhanced by phonons, with selection rules that enable distinguishing neighboring domains. Our multi-technique approach, combining linear optics, TEM, and SFG microscopy, not only demonstrates the presence of DWs, but also enables an estimation of their width. The lateral modulation of in-plane optical anisotropy, as observed in these materials, suggests new avenues for birefringent waveguiding, polariton steering, and frequency conversion, with potential applications in vdW-based photonics and optoelectronics.

\section{Results} 
\label{RnD}

\begin{figure*}[ht!] 
    \centering
    \includegraphics{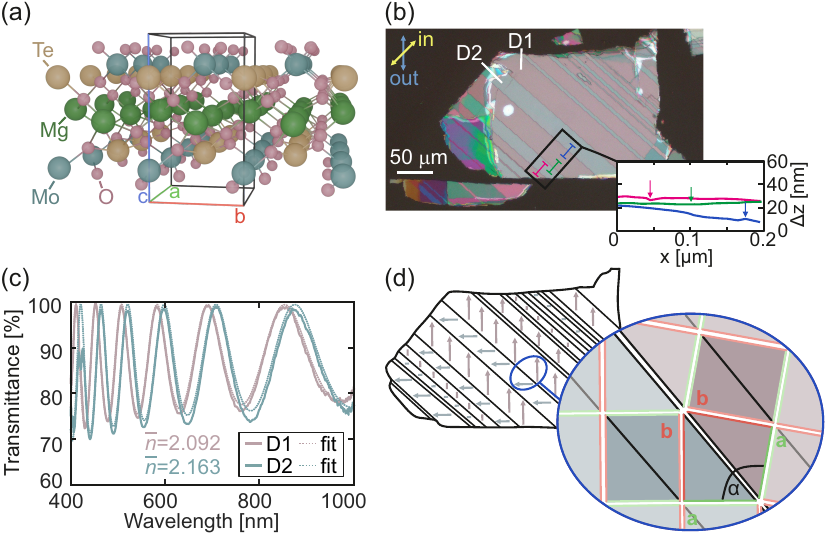}
    \caption{\textbf{Domain observation in \ce{MgTeMoO6}} \textbf{(a)} 2D lattice of \ce{MgTeMoO6}, \textbf{(b)} Polarized optical microscopy image of a \ce{MgTeMoO6} flake with clearly visible stripe domains. The incident/outgoing light is polarized parallel to the yellow/blue arrow, respectively. The image was acquired in reflection geometry. The inset shows atomic force microscopy measurements confirming overall topographic flatness. In some cases, protrusions with $<$\SI{5}{nm} height variations $\Delta z$ - significantly smaller than the flake thickness of \SI{\sim830}{nm} - are observed at the domain walls as highlighted with arrows. \textbf{(c)} Polarized transmittance spectra of two neighboring domains, allowing to extract the flake thickness and the two in-plane refractive indices. For both measurements, light was polarized $\approx$45\textdegree to the DW and, thus, aligned with opposite crystal axes in the two domains. Their spectral averages are $\bar n=$ 2.092 and 2.163, respectively. \textbf{(d)} Hypothesis for the domain wall geometry.}
    \label{fig:fig1}
\end{figure*}

Van der Waals quaternary oxides with the formula \ce{ATeMoO6} with  A = Mg, Mn, Co or Zn have an orthorhombic crystal structure. Figure~\ref{fig:fig1}a shows the orthorhombic unit cell of \ce{MgTeMoO6} embedded in its layered crystal structure. Due to the orthorhombic space group $P2_12_12$, the crystal structure lacks inversion symmetry \cite{O1_botto1980x,O2_forzatti1980metal}. Experimentally, quaternary oxides are grown in bulk. As is typical for vdW crystals, exfoliation yields thin flakes with a well-defined (001) surface that is perpendicular to the \textbf{\textit{c}}-axis. Furthermore, the orthorhombic crystal structure leads to a pronounced in-plane birefringence, similar to other orthorhombic vdW crystals such as \ce{MoO3} \cite{enders2024MoO3}, \ce{V2O5} \cite{le2022V2O5} and \ce{CrSBr} \cite{Ziebel2024CrSBr}, which are actively explored for electrochromic, polarization-sensitive, and magneto-optical device applications.

Optical birefringence can be visualized with POM. A white-light POM reflection image of a large \tsim~\SI{800}{\nm}-thick \ce{MgTeMoO6} flake acquired with polarizer and analyzer offset by 45\textdegree\,is shown in Figure~\ref{fig:fig1}b. There are two distinct features to highlight: First, we observe very pronounced diagonal stripes with alternating colors across the flake, and second, these alternations emerge with different colors in different areas of the flake. The latter observation can be linked to Fabry-Perot resonances that lead to different colors depending on flake thickness \cite{Goldschmidt1984, Manifacier1981}. The diagonal stripes, on the other hand, appear across areas with constant flake thickness, as confirmed by atomic force microscopy (see Figure~\ref{fig:fig1}b inset and SI-Figure 1.1). In this case, the color contrast is rather a result of different in-plane birefringence, suggesting two types of domains. The DWs are aligned parallel to each other throughout the whole flake and even continue across different thickness regions with no offset, suggesting they extend vertically (parallel to \textbf{\textit{c}}) through the entire thickness of the exfoliated flakes. Note that such domains are observed for all four quaternary oxides that share the orthorhombic space group $P2_12_12$, namely \ce{ATeMoO6} with A = Mg, Mn, Zn or Co \cite{S6_tieghi1978crystal}. As A = Cd crystallizes in the tetragonal space group $P\overline{4}2_1m$ neither in-plane birefringence nor domain formation is expected nor observed in this material \cite{O1_botto1980x}. (SI-Figure 2.1 a-e and SI-Figure 2.2 show the respective POM images.) 

To quantify the birefringence and its domain contrast, polarized transmission spectra in the visible to near-infrared spectral range for two neighboring domains are acquired, as shown in Figure~\ref{fig:fig1}c, and used for Sellmeier fits \cite{Weber2003, Sellmeier1871} to extract the refractive index and flake thickness. We obtain an in-plane birefringence of $|n_b - n_a|=|\Delta n| = 0.071$ for \ce{MgTeMoO6}, nearly constant across the visible range. Notably, the same thickness for both domains is extracted from these fits, further confirming the optical rather than topographical origin of the stripe domain contrast. The standing wave appearance of the spectra confirms the colors in POM to be related to a Fabry-Perot cavity effect that depends on the in-plane birefringence \cite{Goldschmidt1984, Manifacier1981}. The fit procedure, the transmission spectra for A = Mg, Mn, Zn, Co and Cd flakes, and the resulting fit values including thickness and refractive indices are reported in SI-Section 2.2.

We observe these stripe domains for all orthorhombic crystals in most flakes across different sizes and thicknesses. For a statistical analysis, we extracted the width of $\sim$500 domains from POM images. SI-Figure 2.1f,g show the histograms for A= Mg and A= Mn, respectively. For A = Mg, more than half of the edges of the analyzed flakes are at 45\textdegree\,to the DW, about a third at 90\textdegree. The stripe domains are a signature of the `top-down' mechanical exfoliation of the vdW materials, often preserving the domain structure of the parent crystal. The fact that domains form naturally in the parent crystal suggests a stability of the DWs that could enable controlled `bottom-up' growth of DWs for potential device applications. Indeed, bottom-up synthesis via chemical vapor deposition has been successfully used to engineer DWs in other layered materials, such as grain boundaries in \ce{MoS2} \cite{van2013grains, najmaei2013vapour} and twin domain alignment in hBN \cite{Wang2019hBN}.

Aside from visualizing the domains, POM helps to further understand their nature. The domains cannot be anti-parallel, as is typical in twin domains in periodically polarized Lithium Niobate, as this would not lead to any contrast in POM, independent of whether the anti-parallel domains are oriented perpendicular or parallel to the surface \cite{myers1995quasi}. Additionally, the domains cannot be ferroelectric as the overall unit cell is apolar \cite{tylczynski2019collection}. Low-temperature antiferromagnetic behavior arising from the magnetic moments of Mn and Co in \ce{MnTeMoO6} \cite{S9_doi2009magnetic} and \ce{CoTeMoO6} \cite{S10_doi2009crystal,S24_Li2025} could indicate domain formation driven by magnetic interactions, as has been hypothesized  for domains observed in \ce{CoTeMoO6} \cite{S24_Li2025}. However, we also find domains with non-magnetic constituents A = Mg or Zn. Hence, the observed domains must be structural in nature with almost perpendicular crystal orientations.

In the POM image (Figure~\ref{fig:fig1}b), the DWs extend without kinks or bends over hundreds of micrometers, which suggests a microscopic geometry that stabilizes the domains and DWs. We thus expect a DW geometry that is lattice-matched to both neighboring domains to enable translational symmetry along the DW. A hypothesis for such a lattice-matched DW structure is shown in Figure~\ref{fig:fig1}d, markedly supporting translational symmetry operations along the domain wall $\textbf{\textit{T}}^{\pm} = n$\textbf{\textit{a}}$ \pm n$\textbf{\textit{b}}$, n \in \mathbb{Z}$, for the two adjacent domains. Using the lattice constants \textit{a}\,=\,5.0378\,\si{\angstrom} and \textit{b}\,=\,5.2691\,\si{\angstrom} \cite{O4_zhang2012mgtemoo}, we expect an angle $\alpha = 92.6^\circ$ between the unit cell vectors \textbf{\textit{a}} in neighboring domains, rather than 90\textdegree. In the following, we will provide experimental evidence supporting this hypothesis, and  will discuss the influence of this domain and DW geometry on the linear and nonlinear optical response.

\begin{figure*}[ht!] 
    \centering
    \includegraphics{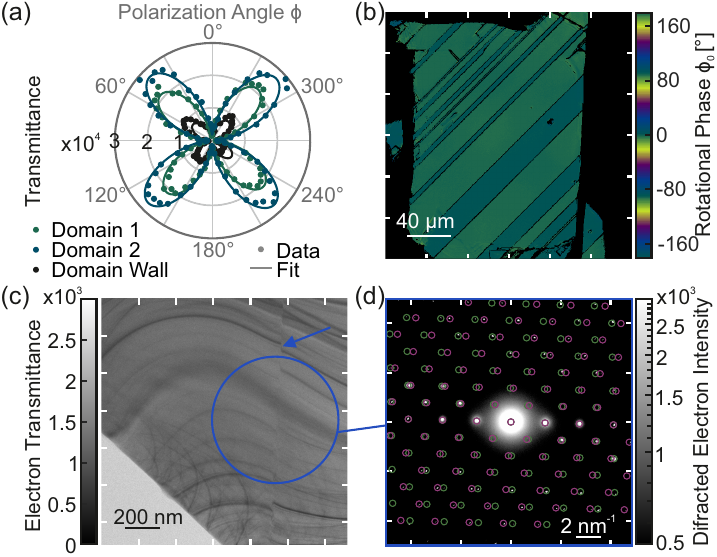}
    \caption{\textbf{Optical and structural domain characterization: }\textbf{(a)} Polarization dependence of the transmitted white light intensity through two neighboring domains of \ce{MgTeMoO6} under crossed polarization and \textbf{(b)} rotational phase $\phi_0$ image, from fits of the polarization dependence with Equation\,(\ref{eq:T_POM}) for each image pixel. \textbf{(c)} TEM image of domain wall of a thin \ce{MgTeMoO6}-flake on a TEM grid. A blue arrow guides the eye to the domain wall. \textbf{(d)} Selected area (blue circle in (c)) electron diffraction pattern acquired at the domain wall. Pink and green circles indicate expected diffraction peak positions for two domains, respectively, with $\alpha = 92.6^\circ$.}
    \label{fig:fig2}
\end{figure*}

Notably, the crossed-polarizer transmission through a in-plane birefringent crystal will be invariant under 90\textdegree~rotation such that for $\alpha\,=$~90\textdegree~no contrast would be expected in a polarization microscope without phase-resolution. Therefore, the observed domain contrast in POM already qualitatively shows that $\alpha$ must deviate from 90\textdegree. Exploiting the full capabilities of POM, we acquired a series of POM images under continuous rotation of both polarizers, being always in the crossed configuration (see SI-Figure~2.6). The intensity of the transmitted light oscillates between bright and dark across the entire flake as the polarizations are rotated. The angle of maximum brightness, however, is offset between neighboring domains, resulting in domain contrast. This domain contrast reverses upon rotation. The experimental setup and exemplary images for A = Mg, Mn, Zn and Co are shown in SI-Section 2.3. Figure~\ref{fig:fig2}a shows the average intensity for two different domains as a function of the polarizer angle. For orthorhombic crystals, the birefringence and, consequently, crossed-polarizer transmission,  $T_{\mathrm{POM}}$, is maximal when the incidence polarization is aligned at 45\textdegree\, to both crystal axes, and vanishes for polarization parallel to the crystal axes, leading to an overall four-fold symmetric signal of the form:
\begin{equation}
    T_{POM} = A \sin^2{[2(\phi + \phi_0)]},
    \label{eq:T_POM}
\end{equation} 
where $\phi$ is the polarizer angle and $\phi_0$ a constant offset. In the following the angle $\phi_0$ is called the `rotational phase' - it is a quantitative measure for the optical anisotropy. The four-fold symmetry predicted by Equation\,(\ref{eq:T_POM}) is observed experimentally in Figure~\ref{fig:fig2}a. Extending this analysis to every pixel and fitting using Equation\,(\ref{eq:T_POM}), the rotational phase can be extracted across the image as shown in Figure~\ref{fig:fig2}b. A bimodal phase distribution shows a clear alternation between two types of domains. We extract an offset of $\Delta\phi_0 = 3.2 \pm0.7$\textdegree\ between the two types.  Accounting for the four-fold symmetry, this offset corresponds to  $\alpha = 93.2\pm0.7$\textdegree\,, in good agreement with the hypothesis in Figure~\ref{fig:fig1}b predicting  $\alpha = 92.6$\textdegree. Polar plots and rotational phase maps for A = Mg, Mn, Zn and Co, as well as details on the fitting procedure are provided in SI-Section 2.3.

Linearly-polarized optical microscopy efficiently resolves the optical anisotropy of each domain and its difference between the domains. However, the hypothesis in Figure~\ref{fig:fig1}d suggests an atomistic DW structure on the nanometer scale. In order to connect the optical results with a probe of the crystal structure itself, we performed TEM measurements of a $\sim$100~nm thin flake of \ce{MgTeMoO6} that was placed on top of a \ce{SiN_x} support grid with apertures as shown in SI-Figure 3.1. Figure~\ref{fig:fig2}c shows a wide-field TEM image of a region that contains a domain wall. The DW is clearly visible due to the abrupt offset of wavelike patterns at the DW. These patterns likely result from electron interference due to slight bending/tilting and associated mild strain across the suspended flake \cite{Fultz2013}. Fortunately, the offset in the strain features provides direct DW visibility, which also allowed for selecting sample areas for electron diffraction exclusively within each single domain or across a DW including both domains. Figure~\ref{fig:fig2}d shows a selected area electron diffraction pattern for the latter case. The corresponding sample area including both domains is labeled in ~Figure~\ref{fig:fig2}c. In the diffraction pattern, each Bragg peak position shows horizontally displaced double spots, with their distance increasing away from the central zero-order peak. Diffraction patterns recorded at either side of the DW instead only yield either one of the double spots. When simulating a structure by overlaying the reciprocal lattice from two rotated orthorhombic lattices to match the experimental diffraction pattern, it was found that only a relative rotation of 87.43\textdegree\ (or equivalently 92.57\textdegree) produces a suitable match, as visualized by the green and violet reciprocal lattices, hence perfectly matching the hypothesis. The  reciprocal lattice calculation is detailed in SI-Section 3.2. To summarize, both the optical anisotropy using POM and the structural analysis using TEM support the hypothesis for the DW structure shown in Figure~\ref{fig:fig1}d and the resulting prediction for rotation angle $\alpha$ between the domains.

\begin{figure*}[ht] 
    \centering
    \includegraphics{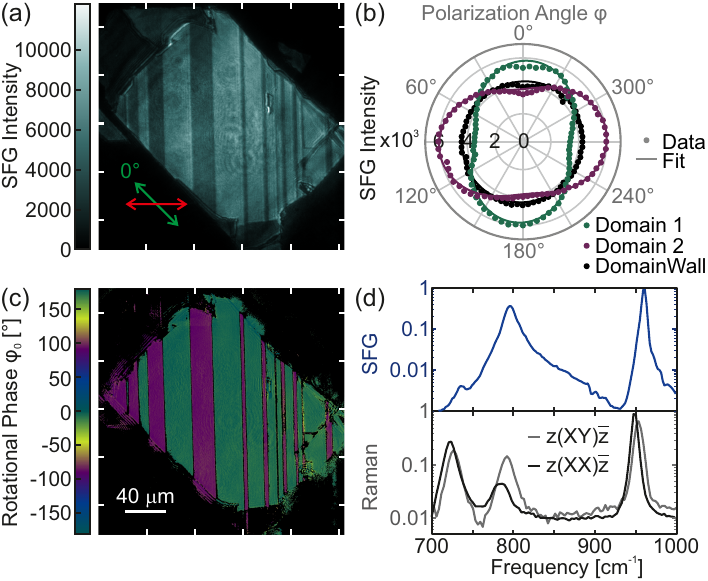}
    \caption{\textbf{Second-order nonlinear optical response of \ce{MgTeMoO6}}: \textbf{(a)} Real space SFG image at $\omega_\text{IR}= 961\,{\wn}$ with clear domain contrast. The in-plane laser polarization components of the VIS and IR are shown in green and red, respectively. \textbf{(b)} Domain-averaged SFG intensity of neighboring domains, and for the domain wall as a function of VIS laser polarization angle $\varphi$. \textbf{(c)} Rotational phase $\varphi_0$ image extracted from fits of the polarization dependence with Equation\,(\ref{eq:SFG_model}) for each image pixel. \textbf{(d)} Top: Normalized SFG-spectrum as function of IR-laser frequency in blue. Bottom: Normalized Raman spectra for crossed and parallel configuration in gray and black, respectively.}
    \label{fig:fig3}
\end{figure*}

The strongly anisotropic response of the QO in linear optics is promising for technological applications. Beyond this, the QO also exhibit large second-order nonlinearities, broadening the application range to nonlinear optics. To understand how the abrupt changes in in-plane anisotropy between neighboring domains carry over into the nonlinear optical response, we employ IR-VIS SFG spectro-microscopy \cite{mader2024sum,niemann2022long}. In this method, a narrow-band IR laser beam is upconverted with a VIS laser beam, enabling resonant IR imaging with a spatial resolution determined by the wavelength of the VIS upconversion laser. Figure~\ref{fig:fig3}a shows an SFG image at $\omega_\text{IR}= 961\,{\wn}$, clearly showing an intensity contrast between neighboring domains. Upon rotation of the polarization of the VIS beam, the domain contrast inverts, see SI-Figure 4.3. Being a $\chi^{(2)}$-process, the SFG is sensitive to lattice symmetry. Therefore, the SFG intensity depends on the experimentally chosen laser polarizations, which must align with the symmetry-allowed tensor elements of $\chi^{(2)}$. For the orthorhombic QOs and the illumination configuration employed here, the SFG intensity is expected to follow a two-fold symmetry when rotating the sample or, complementary, the polarization of the VIS laser. When rotating the VIS polarization by angle $\varphi$, and using p-polarized IR illumination (fixed), the SFG intensity is expected to follow
\begin{equation} 
I_\mathrm{SFG}(\varphi) = |B\cdot \sin(\varphi+\varphi_0)|^2+|C\cdot \cos(\varphi+\varphi_0)|^2,
\label{eq:SFG_model}
\end{equation}
where $B$ and $C$ incorporate $\chi^{(2)}_{abc}$ and $\chi^{(2)}_{bac}$, respectively, including angle-independent local field constants. A detailed derivation can be found in SI-Section 4.1.

Indeed, the angular dependence of the experimental SFG intensity for each domain follows a two-lobed shape, which can be seen in the domain-integrated SFG intensity plotted in Figure~\ref{fig:fig3}b. Using Equation\,(\ref{eq:SFG_model}), the rotational phase difference, $\Delta\varphi_0$, between neighboring domains is found to be 100$\pm$4\textdegree, which - despite qualitatively matching the DW hypothesis - deviates from the expected 92.6\textdegree. This is likely related to linear-optical effects, such as birefringence and Fabry-Perot resonances, affecting the SFG azimuthal response. Yet, similarly to the crossed-polarizer POM data, also for SFG, a rotational phase $\varphi_0$ can be extracted for each pixel, see Figure~\ref{fig:fig3}c. Its bimodal distribution again confirms that only two types of domains exist. Note that opposed to the linear birefringence, which is expected to peak inbetween crystal axes, SFG is expected to peak when the VIS laser polarization is aligned with the crystal axis associated with the larger $\chi^{(2)}$-element in Equation\,(\ref{eq:SFG_model}), i.e.\,$|\chi^{(2)}_{abc}|$ or $|\chi^{(2)}_{bac}|$.

The dependence of the SFG intensity on the VIS polarization angle shown in Figure~\ref{fig:fig3}a-c was measured with an IR laser frequency $\omega_{\mathrm{IR}}= 961\,{\wn}$ that matches a phonon resonance. Imaging at a different resonance at $\omega_{\mathrm{IR}}=796\,{\wn}$ (not shown) leads to comparable rotational phase and polarization dependence. The IR-frequency-dependent SFG spectrum for \ce{MgTeMoO6} is shown in Figure~\ref{fig:fig3}d, where both resonances can be clearly distinguished. As SFG signals will get enhanced from phonons being simultaneously IR- and Raman-active \cite{liu2008sum,mader2024sum}, Raman spectra are also plotted for comparison, showing that the SFG intensity maxima indeed coincide with Raman peaks of \ce{MgTeMoO6}. The phonon resonances lead to an increase of the SFG intensity by up to four orders of magnitude, enabling rapid nonlinear imaging of the QOs. A detailed description of the underlying SFG theory, a depiction of the setup, SFG images, polarization dependence, and rotational phase maps for A = Mg, Mn, Zn, Co, and Cd can be found in SI-Section 4. In general, the nonlinear optical response confirms the domain geometry proposed in this work,  and could, for instance, enable applications for frequency conversion modulated on the nanoscale between twin domains.

\begin{figure*}[ht] 
    \centering
    \includegraphics{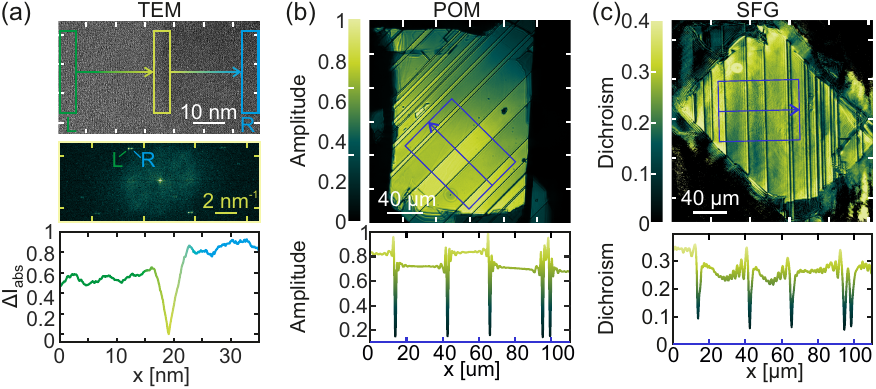}
    \caption{\textbf{Domain wall analysis for \ce{MgTeMoO6}}: \textbf{(a)} Real space TEM image (top) with Fourier transformation (middle) at domain wall (yellow box). Normalized intensity difference between the left (green L) and right (blue R) side diffraction peak $\Delta I_{\text{abs.}}$ as function of box position (bottom). \textbf{(b)} Four-fold modulation amplitude of rotational crossed-polarizer microscopy (top) with line cut over blue box (bottom). \textbf{(c)} SFG dichroism $\mathcal{D}$ image (top) with line cut over blue box (bottom).}
    \label{fig:fig4}
\end{figure*}

In our hypothesis in Figure~\ref{fig:fig1}d, the DWs act as a mirror plane, leading to a prediction for the angle $\alpha$ between crystal axes of neighboring domains, as confirmed by POM and TEM measurements shown in Figure~\ref{fig:fig2}. Our proposed DW structure, however, would confine the DW itself to only the one unit cell located directly at the domain interface. To experimentally explore this scenario, the wide-field TEM data (Figure~\ref{fig:fig2}c) first provide a clear visual estimate of the DW width to be below \SI{30}{\nm}. This upper limit is refined to below \tsim~\SI{5}{\nm} using a higher-magnification image shown in the top panel of Figure~\ref{fig:fig4}a. Despite not being easily visible by eye, the wide-field TEM image resolves the lattice periodicity sufficiently to enable a more analytical approach to estimate the DW width: A sliding-window ($5.2 \times 13.1$~nm)
two-dimensional Fourier transformation, as indicated in Figure~\ref{fig:fig4}a, allows us to trace the domain-specific diffraction spots (middle panel) when sliding perpendicular to the DW. The bottom panel shows the normalized intensity difference $\Delta I_{\text{abs.}} = |(I_L-I_R)/(I_L+I_R)|$. The DW width can be extracted as the full-width-at-half-maximum (FWHM) of the dip in $\Delta I_{\text{abs.}}$ at the DW, which was found to be $4.7 \pm 1.9$~nm. Wide-field imaging with such a relatively thick flake constrains the achievable TEM resolution to about this value, such that the extracted DW width only provides an upper limit for its physical width, see SI-Section~3.3 for details. The TEM data and analysis does not provide atomically resolved evidence of the proposed DW structure at the current resolution, yet helps to exclude any structures that would extend beyond \SI{5}{nm}.

The POM and SFG microscopy data discussed in Figures~\ref{fig:fig2} and \ref{fig:fig3} can also specifically provide DW contrast, which is shown in Figure~\ref{fig:fig4}b,c. In these methods, the spatial resolution is \tsim~\SI{1}{\um}, as determined by the diffraction limit, and is thus clearly worse than in TEM. Yet, it is interesting to explore DW contrast mechanisms in these optical methods. Note that any DW contrast observed by optical means is unlikely to arise directly from the DW region itself, due to the small DW width of \SI{<4.7}{nm} (according to TEM) compared to the \SI{\sim 1}{um} spatial resolution of the optical techniques. Instead, the DW contrast is expected to emerge from spatial averaging of signals from the much larger volumes of the two domains on either side of the DW. 

For POM, the signal magnitude drops at the DWs, as seen in the angle dependence of the DW shown as black dots in Figure~\ref{fig:fig2}a. This suggests that the opposite birefringence from the adjacent domains compensates, minimizing the cross-polarized transmission, such that the four-fold modulation amplitude vanishes. For SFG, the VIS polarization angle dependence at the DWs shown as black points in Figure~\ref{fig:fig3}b takes a circular shape, while the average SFG signal magnitude is largely preserved. Following these observations, we specifically plot the POM modulation amplitude ($A$ in Equation\,(\ref{eq:T_POM})) in Figure~\ref{fig:fig4}b and the SFG dichroism, $\mathcal{D}$, in Figure~\ref{fig:fig4}c, each shown as maps (top panel) and as line-cuts  perpendicular to the DWs (bottom panel). The SFG dichroism is defined as $\mathcal{D}= |B^2-C^2|/(B^2+C^2)$, with $B,C$ being the signal amplitudes along the symmetry axes, see Equation\,(\ref{eq:SFG_model}) and SI-Section~4 for details. These parameters maximize DW contrast to help analyze and trace DWs specifically using optical methods. An equivalent analysis of A = Mn, Zn, and Co for POM and SFG is provided in SI-Section 2.4 and SI-Section 4.4, respectively. Overall, wide-field POM and SFG imaging provide not only domain- but also DW contrast, thereby enabling rapid screening for suitable sample areas for any given future application. 

\section{Discussion and Outlook}
This study presents a comprehensive investigation of naturally occurring in-plane twin domains in orthorhombic van der Waals quaternary oxides \ce{ATeMoO6} (A = Mg, Mn, Co, Zn).  The DWs act as diagonal mirror planes, providing alternating linear optical birefringence and domain contrast across a single flake without external patterning. Similar domain contrast is observed for the  second-order nonlinear response, which further shows pronounced enhancement at phonon resonances \cite{urban2026thz,mueller2026full}. Future work will investigate the spectroscopic behavior in SFG, systematically mapping the nonlinear response across the mid-IR phonon spectrum to explore the role of phonon resonances and symmetry breaking in the twin-domain system. This will further establish QOs as a platform for phonon-engineered nonlinear nanophotonics.

Furthermore, to understand the influence of the A-site cation on birefringent and nonlinear properties, we systematically compared \ce{ATeMoO6} (A = Mg, Mn, Zn, Co, and Cd) in SI-Section 7. The ionic radius primarily dictates the orthorhombic unit cell parameters (\textbf{\textit{a}}, \textbf{\textit{b}}) and $\alpha$ with larger cations (e.g. Cd) driving towards tetragonal symmetry ($\alpha = 0^\circ$). However, the optical anisotropies — linear birefringence $\Delta n$ and SFG dichroism $\mathcal{D}$ — show no clear correlation with ionic radius and, intriguingly, are anti-correlated. Note the clear trend between unit cell parameters and ionic radii (SI-Fig.~7.1a)  supporting the ionic character of A as \ce{A^{2+}}; however, for readability purposes we chose to label them as A throughout the text. 

Our combination of techniques enabled the characterization of the domain structure and an upper-bound estimation of the DW width. An important question is whether these domains can also be manipulated or even engineered. In that regard,  an intriguing  observation was made unintentionally: When one of the samples was dropped, a piece of a \ce{MgTeMoO6} flake broke off, while also some domains switched and new domains formed within parts of the remaining flake, see SI-Figure~2.3 for a direct comparison. This observation suggests that the crystal could be ferroelastic. In principle, ferroelasticity is allowed in the orthorhombic space group $P2_12_12$ with {110}-type domain walls \cite{aizu1970possible}. Given that the nearly iso-structural Cd-compound crystallizes in a closely related tetragonal space group, a force-induced transition via this higher-symmetry state is conceivable. 
First-principles calculations predict an energy barrier of $\approx$0.25~eV per unit cell~(14~meV/atom) for the ferroelastic phase transition in \ce{MgTeMoO6}, well in the range of typical values for ferroelastic switching \cite{carbogno2014ferroelastic,Xuan.2022}, also see SI-Section~6. With such a low energy barrier, it should be feasible to actively manipulate the domains, for instance by applying uniaxial stress \cite{roeper2026uniaxialstraindrivenferroelasticdomain} or nano-indentation \cite{gao2014ferroelastic}. If proven successful, potential applications of orthorhombic QOs would further broaden to not just engineer optical structures but further to also use the materials for actuators, sensors, and reconfigurable domain architectures in industrial applications.

Our work reveals the functional significance of domains and DWs in optical control: the abrupt change in in-plane anisotropy enables polarization-dependent frequency conversion, hyperbolicity switching, and the emergence of domain-specific nonlinear signals. These naturally formed domains offer a compelling, low-cost route to engineer lateral optical interfaces in van der Waals heterostructures bypassing the need for complex nanofabrication. Future efforts should focus on controlling the domain structures, either by  nucleation and orientation during crystal growth or by optimizing stress-induced domain manipulation \cite{roeper2026uniaxialstraindrivenferroelasticdomain}, or even by resonant phonon excitation \cite{kwaaitaal2024epsilon}.  Such domain control would enable the targeted design of domain patterns for applications in nanoscale polarization optics such as zone plates, polariton waveguiding, and nonlinear metasurfaces using pseudo-phase matching or nonlinear holographic imaging. The integration of \ce{ATeMoO6} with other 2D materials could further exploit interfacial symmetry breaking for enhanced functionality.

\section{Conclusion}
In this work we reported the observation of ubiquitous twin domains in a whole family of orthorhombic van der Waals quaternary oxides \ce{ATeMoO6} (A = Mg, Mn, Zn, Co). A combination of linear polarization microscopy, nonlinear SFG microscopy, and structural characterization based on TEM demonstrates that the crystal structure and DW geometry are responsible for the domain properties. We show that the crystal rotation between domains is dictated by the dimensions of the orthorhombic unit cell through the ratio of the in-plane lattice constants. Pronounced domain contrast of linear birefringence and SFG anisotropy provides a rich platform for ultrathin polarization optics. Preliminary data suggest that domains can also be manipulated, indicative of ferroelastic behavior, which would further enhance potential for applications. Overall, twin domains in low-symmetry quaternary oxide van der Waals flakes reported here will provide many opportunities for applications for nanoscale polarization optics, polariton engineering, and metasurfaces using vertical and lateral heterostructures.

\section*{Supporting Information}
The supporting information shows detailed data and fitting procedures for the different techniques. Section 1 shows the topography across domain walls measured by atomic-force microcopy. Section 2 outlines linear optics including POM, transmission spectroscopy, birefringence and domain wall analysis for all four materials, including a description of the underlying physics and fitting procedures. Section 3 shows the simulation of diffraction peaks and domain wall analysis in TEM. Section 4 shows the SFG theory including a sketch of the setup, a complementary sample rotation, estimation of $\chi^{(2)}$ and the analysis of the angular rotation and domain wall fitting for all four materials plus the reference material \ce{CdTeMoO6}. Section 5 provides more extensive Raman spectroscopy data. Section 6 discusses the first-principles calculations that elucidate and give evidence for ferroelastic behavior of \ce{MgTeMo6}. Section 7 shows a comparison of the different materials \ce{ATeMo6} and discusses the resulting properties in relation to the atomic radii of A. 

\section*{Methods}
\subsection*{Sample preparation}
Single crystals of \ce{ATeMoO6} (A~=~Zn, Mn, Co, Mg, Cd) were grown using a self-serve melt method as described elsewhere \cite{O17_sun2024van, O6_zhao2013zntemoo, O8_zhao2013combination}. Subsequently, a large selection of flakes was mechanically exfoliated onto 1~mm-thick \ce{CaF2} substrates for each material using conventional scotch tape exfoliation.

\subsection*{Polarized Optical Microscopy}
For the linear optical microscopy image in Figure~\ref{fig:fig1}b and SI-Section 2.1 an Olympus BX51M microscope in reflection mode was used including MPlanFL N 5x, 10x, 20x and 50x objectives, an Olympus TH4-200 halogen lamp, a light adjustment filter U-25LBD, an incoming polarizer U-PO3, a rotatable analyzer U-AN360-3 and a 5 megapixel digital camera DP26.

Polarized optical microscopy (POM) with crossed polarizers was performed in transmission geometry using a home-built setup, schematically shown in SI-Section 2.3. A broadband thermal white-light source illuminated a linear polarizer, producing a well-defined linearly polarized beam. A condenser lens focused the light onto the sample, which consisted of quaternary oxide flakes on a \ce{CaF2} substrate. The transmitted light was collected by a 100x microscope objective (Olympus LMPlanFl) and passed through a second linear polarizer (analyzer) before being recorded by a CMOS camera (Retiga E9, Teledyne). Both, the polarizer and analyzer were automated and could be rotated independently, allowing measurements under orthogonal polarizer-analyzer configuration. The retardation introduced by the sample converted the incident linear polarization into an elliptical state, which the analyzer then mapped into intensity contrast on the camera.

\subsection*{Polarization-Resolved Transmission Spectroscopy}
The transmission spectra in Figure~\ref{fig:fig1}c and SI-Section 2.2 were recorded using a home built inverted bright field microscope \cite{Juergensen2024}. As an excitation source a supercontinuum laser (NKT FIU15) with a broad spectral range (390--2400\,nm) was used.
The broadband laser light consisting of all wavelengths was guided to a microscope (Olympus- IX71) equipped with a 100× objective (NA 0.9) that focused the light onto the sample, and a second 100× objective (NA 0.8) to collect the transmitted light. Via an optical fiber the collected light was guided to the spectrometer (Avantes, AvaSpec ULS2048CL-EVO). A wire-grid polarizer mounted on an automated rotation stage in the beam path ensured fully linearly polarized light, which was used to probe the sample orientation.

\subsection*{Transmission Electron Microscopy}
For the TEM measurement the flakes were transferred onto a hole in a Norcada \ce{SiN_x} membrane (thickness \SI{200}{\nm}, window size 50 x 50 \SI{}{\um}, hole diameter \SI{5}{\um}, hole array size 4x4) using a dry pick-up technique with a PDMS/PPC stamp \cite{pizzocchero2016hot, kim2016van}. 
TEM imaging was performed on a ThermoFisher Titan 80-300 operated at \SI{300}{\kV} and equipped with a TVIPS XF416 CMOS camera. The sample was mounted on a double-tilt sample holder. An objective aperture (\SI{70}{\um}) was used to improve the contrast of the TEM images. The thickness of the \ce{MgTeMoO6} flake imaged in Fig.~2 and 4 was measured to be $\approx$\SI{100}{nm} using AFM before transfer onto the TEM hole array, and the flake was apparently to thick for achieving atomic resolution in the TEM. Mechanical transfer of thinner flakes with pre-characterized domains was not successful. 

\subsection*{SFG Spectro-Microscopy}
The experimental details of the SFG microscope were previously described in detail \cite{mader2024sum, niemann2024spectroscopic, niemann2022long, niemann2026vectorfield}. In the present work, a frequency-doubled amplified fiber laser ($\lambda_{\text{VIS}}$~=~\SI{520}{\nm}, Orange High Power 10, Menlo, 1040 nm, repetition rate 55.5~MHz, pulse energy 10~nJ) was spatially and temporally overlapped with the FHI-FEL (oblique incident angle $\theta$~=~50\textdegree, \SIrange{650}{1050}{\wn}, repetition rate 55.5 MHz). The filtered SFG beam was detected using a gated CCD camera (1024x1024 pixel, Teledyne PI-Max-4) with a 50x objective (Mitutoyo M Plan Apo 50×, NA\,=\,0.55, working distance wd\,=\,13\,mm). The resulting field of view was $275 \times 275$\,\SI{}{\um}. Other experimental conditions such as details on the pulse structure of the FEL and its synchronization \cite{kiessling2018femtosecond,schollkopf2015new} and details on the focusing and detection scheme \cite{niemann2022long,niemann2026vectorfield} can be found in the respective references.

\subsection*{Atomic Force Microscopy}
Atomic force microscopy (AFM) topography measurements were performed on either an AFM system from attocube GmbH (neaSCOPE, Munich, Germany) using a Pt-coated Si AFM tip (Arrow-NCPt, NanoWorld) or a Park Systems AFM (XE-150) using a silicon tip (PPP-NCHR, Nanosensors). Measurements were conducted in non-contact mode with tapping frequencies of $\sim$330 kHz or $\sim$285 kHz respectively.

\subsection*{Raman Spectroscopy}
To measure polarized Raman spectra, a Raman microscope (Horiba, XploRA) in reflective configuration was used. It consists of a standard bright field microscope (Olympus BX51) with exchangeable objectives and an adjustable $xyz$-piezostage. The sample was mounted on a manually operated rotation stage, allowing to selectively choose sample orientation. All spectra were recorded with 532 nm excitation wavelength and dispersed with a 1200 gr/mm grating onto a CCD camera.

\subsection*{First-Principles Modeling}
To shed light on the viability of a ferroelastic switch in \ce{MgTeMoO6}, we performed density-functional theory calculations with the \textit{FHI-aims} package~\cite{Blum.2009,Abbott.2026} at the semi-local level of theory. In particular, we determined the minimum-energy path between two symmetry-equivalent $P2_12_12$ structures with reoriented lattice vectors~$a,b$ via the generalized solid-state nudged elastic band approach~\cite{Sheppard.2012}. More details are given in Sec.~6 of the SI.

\subsection*{AI assistance}
In this study, various AI models were consulted for advisory support on topics including language formalism, conventions, coding (specifically for LaTeX and MATLAB), and mathematical proof validation. The primary models used included Claude (Anthropic), Perplexity, and—within the ChatAI interface—Devstral 2 123B Instruct 2512, Gwen, and Mistral. While these were the most frequently employed models, others were occasionally utilized as needed.

\section*{Acknowledgements}
D.S.M acknowledges funding from the Max Planck-Radboud University Center for Infrared Free-Electron Laser Spectroscopy. N.S.M. acknowledges funding from the Deutsche Forschungsgemeinschaft (DFG, German Research Foundation) - Projektnummer 551280726.  S.F.M. acknowledges funding from the Deutsche Forschungsgemeinschaft (DFG, grant number 469405347). N.S. acknowledges funding through the Collaborative Research Center CRC1772 (project B01) and the German Excellence Strategy - EXC3112/1 - 533767171 (Center for Chiral Electronics). X.Y and P.L acknowledge funding from National Key Research and Development Program of China (2024YFA1208500), National Natural Science Foundation of China (62525504, 62505104), Scientific Research Innovation Capability Support Project for Young Faculty (ZYGXQNJSKYCXNLZCXM-I17) and the Hubei Optical Fundamental Research Center (HBO2025TQ005). We thank S. Zhao (Chinese Academy of Sciences Fuzhou) for kindly supplying the \ce{ATeMoO6} crystals. We thank Prakriti P. Joshi (Stony Brook University) for providing code for pixel-wise data analysis. We thank Stephanie Reich (FU Berlin) for providing access to the Raman spectroscopy and micro-transmission setups.

\section*{Conflicts of Interest}
The authors have no conflicts to disclose.

\section*{Data Availability}
The data that support the findings of this study are openly available in Zenodo at https://doi.org/10.5281/zenodo.22710429.

\end{multicols}

\clearpage

\end{document}